\documentclass[a4paper]{IEEEtran}

\usepackage{booktabs}
\usepackage{amsmath}
\usepackage{amsthm}
\usepackage{multirow}
\usepackage{color,cite}
\usepackage{subfigure}
\usepackage{mathrsfs}
\usepackage{amssymb}
\usepackage{bm}
\usepackage{graphicx}
\usepackage{amsfonts}
\usepackage{algorithm}
\usepackage{algorithmic}
\usepackage{epstopdf}

\begin{document}

\bibliographystyle{IEEEtran}

\title{\vspace{-10mm}\Large{HyperCEUNet: Parameter-Aware Hypernetwork-Driven UNet for Channel Estimation}
}

\author{\vspace{-2mm}
Ke Ma, Feng Wang, Lihui Lei, Shu Tan
    \thanks{{\em Corresponding author: Feng Wang.}}
    \thanks{Ke Ma, Feng Wang, Lihui Lei and Shu Tan are with XringTek, Xiaomi Inc., Beijing 100085, China (e-mail: make1@xiaomi.com; wangfeng\_1989@126.com; leilihui@xiaomi.com; tanshu@xiaomi.com).}

    \vspace{-12mm}
}

\maketitle

\begin{abstract}

Deep learning-based channel estimation has been recognized as a promising technique for sixth-generation wireless systems. However, most existing approaches rely solely on least-squares estimates obtained from demodulation reference signals, which fail to explicitly exploit channel time-frequency correlation parameters. Inspired by the independent channel parameter estimation enabled by semi-static reference signals in modern wireless systems, this letter presents a parameter-aware deep learning-based channel estimation framework termed HyperCEUNet. Specifically, the proposed hypernetwork generates an adaptive front-end convolutional layer based on estimated channel parameters, serving as a pre-filtering stage before the UNet-based estimator. In addition, the Wiener-filtered channel estimates are adopted to provide a correlation-aware initialization for data resources. Simulation results demonstrate that our proposed HyperCEUNet effectively improves channel estimation accuracy compared with its conventional counterparts.

\end{abstract}


\IEEEpeerreviewmaketitle

\vspace{-4mm}

\section{Introduction}\label{S1} 

In modern wireless systems, channel state information (CSI) is generally obtained through pilot-aided channel estimation, where reference signals are embedded into the time-frequency grid and the channel on data resources is inferred from pilot observations \cite{3gpp1}. As one of the fundamental modules, channel estimation plays a critical role in accurately tracking the frequency-selective and time-varying characteristics of wireless channels.
Conventional pilot-based channel estimation methods primarily rely on linear estimators, such as least-squares (LS) and Wiener filter-based linear minimum mean square error (LMMSE) \cite{lmmse}. However, these conventional methods may not achieve satisfactory performance due to their limited capability in modeling complex nonlinear channel characteristics.

With the evolution toward next-generation wireless systems, deep learning-based channel estimation has attracted growing interest due to its ability to learn complex nonlinear mappings in a data-driven manner.
Existing deep learning methods typically formulate this problem as a two-dimensional signal recovery task in the time-frequency domain\cite{channelnet, reesnet, ceunet, attenreesnet, sccenet}. As one of the earliest frameworks, ChannelNet\cite{channelnet} regards channel estimation as an image super-resolution task, and adopts convolutional neural network (CNN) to recover the full-resolution channel. ReEsNet\cite{reesnet} learns the residual between the initial estimate and the ground truth, which improves learning efficiency and mitigates the vanishing gradient issue. Based on this residual framework, AttenReEsNet\cite{attenreesnet} further utilizes the attention mechanism to enrich feature representation. Besides, CRCENet\cite{ceunet} designs a compression–reconstruction network based on UNet\cite{unet} to improve channel estimation performance. Moreover, SCCENet\cite{sccenet} integrates symmetric CNN with attention to implement efficient and robust channel estimation.
{\color{black}Recently, diffusion model-based channel estimation has emerged as a promising direction. For example, the work \cite{diffusion} learns a generative diffusion prior of multiple-input multiple-output (MIMO) channels and incorporates it into a variational inference framework for robust channel estimation.}

Despite the rapid progress in deep learning-based channel estimation, most existing works solely adopt demodulation reference signal (DMRS)-based LS estimates as the model input \cite{channelnet, reesnet, ceunet, attenreesnet, sccenet}. This design inherently overlooks key characteristics of modern wireless systems. Specifically, DMRS shares the same precoding and time-frequency allocation as the associated data. On the one hand, New Radio (NR) allows different precoding across time-domain subframes and frequency-domain bundles \cite{3gpp1}. Consequently, DMRS-based LS estimates may correspond to different effective channels, making their joint utilization challenging. On the other hand, flexible data scheduling may allocate only a limited number of subcarriers in the frequency domain \cite{3gpp2}, resulting in sparse DMRS observations and degraded channel estimation quality.

In contrast, modern wireless systems generally employ semi-static reference signals, such as tracking reference signal (TRS) in NR, to estimate channel parameters including Doppler spread, delay spread, and signal-to-noise ratio (SNR) \cite{3gpp1}. Since these semi-static reference signals are not subject to precoding and typically span the entire system bandwidth, they can accurately characterize channel time-frequency correlations and enable effective Wiener filtering. Nevertheless, most deep learning-based channel estimation approaches do not leverage such parameter information \cite{channelnet, reesnet, ceunet, attenreesnet, sccenet}. A few recent studies \cite{joint1, joint2} have considered joint parameter-channel estimation. However, their focus is mainly on extracting channel parameters from  reference signals used for channel estimation, and their scope is limited to carrier frequency offset compensation.

To address the above limitation, this letter proposes a parameter-aware deep learning-based channel estimation framework termed HyperCEUNet. {\color{black}Different from existing DMRS-only deep learning-based methods, the key idea of this work is to explicitly incorporate TRS-assisted channel parameters into the estimator through a lightweight hypernetwork-driven adaptation mechanism.} UNet \cite{unet} is adopted as the backbone due to its strong capability in capturing multi-scale features in the time-frequency domain, and channel attention (CA) \cite{ca} is harnessed to enrich feature representation. To incorporate channel parameter information, a hypernetwork is introduced to generate an adaptive front-end convolutional layer conditioned on estimated channel parameters, including time-domain selectivity, frequency-domain selectivity, and SNR. This layer performs pre-filtering before UNet, serving as a data-driven and parameter-conditioned generalization of the Wiener filter.
Besides, the Wiener-filtered channel estimates are utilized as a correlation-aware initialization for the data resource grid, augmenting prior information that reflects the time-frequency channel statistics from parameter estimation.
Simulation results show that our proposed HyperCEUNet achieves superior estimation accuracy compared with its conventional counterparts.\footnote{In spite of using UNet as the backbone in this letter, our hypernetwork-based parameter incorporation framework can be easily applied to other backbone models.}

\section{System Model}\label{S2} 
Consider a downlink orthogonal frequency-division multiplexing (OFDM) system with $N_{\text{f}}$ subcarriers and $N_{\text{t}}$ OFDM symbols per transmission duration, thus there are $N_{\text{f}}N_{\text{t}}$ resource elements (REs). The received signal on the $k$-th subcarrier and the $n$-th OFDM symbol after removing the cyclic prefix and performing discrete Fourier transform is given by
\begin{equation} \label{eq1}
y(k,n)=h(k,n)x(k,n)+w(k,n),
\end{equation}
where $x(k,n)$ denotes the transmitted symbol, $h(k,n)$ represents the frequency-domain channel, and $w(k,n)$ is additive white Gaussian noise (AWGN) with variance $\sigma_\text{w}^2$.
The channel is assumed to be frequency-selective and time-varying, while remaining constant within one RE.

DMRSs are inserted into the OFDM time-frequency resource grid according to a predefined pattern compliant with NR standards. Let $\Omega \triangleq \{(k_{\text{p}}[i], n_{\text{p}}[j]) \mid i=1,\cdots,N_{\text{f,p}},\, j=1,\cdots,N_{\text{t,p}}\}$ denote the set of pilot positions, where $N_\text{f,p}$ and $N_\text{t,p}$ correspond to the number of pilots in the frequency and time domains, respectively. The received signal on DMRS locations is given by
\begin{equation} \label{eq2}
y_{\text{p}}(k,n) = h(k,n) x_{\text{p}}(k,n) + w(k,n), \quad (k,n)\in \Omega,
\end{equation}
where $x_{\text{p}}(k,n)$ represents the known DMRS symbol.
Based on these pilot observations, the receiver aims to estimate the full channel matrix $\mathbf{H}\in\mathbb{C}^{N_\text{f}\times N_\text{t}}$, whose $(k,n)$-th entry is $h(k,n)$.

The LS channel estimates obtained on DMRS locations are denoted by $\hat{\mathbf{H}}_{\text{p,LS}} \in \mathbb{C}^{N_\text{f,p} \times N_\text{t,p}}$, and their $(i,j)$-th entry is
\begin{equation} \label{eq3}
\hat{h}_{\text{p,LS}}(i,j) = \frac{y_{\text{p}}(k_{\text{p}}[i],n_{\text{p}}[j])}{x_{\text{p}}(k_{\text{p}}[i],n_{\text{p}}[j])}, \quad (k_{\text{p}}[i],n_{\text{p}}[j]) \in \Omega.
\end{equation}
By further leveraging the second-order channel statistics, the Wiener filter-based LMMSE estimator computes the channel estimates at data positions as
\begin{equation} \label{eq4}
\hat{\mathbf{H}}_{\text{d,wn}} = \text{mat} \left( \mathbf{R}_{\mathbf{h}_\text{d},\mathbf{h}_\text{p}} \left( \mathbf{R}_{\mathbf{h}_\text{p},\mathbf{h}_\text{p}} + {\nu}^{-1} \mathbf{I} \right)^{-1} \text{vec}(\hat{\mathbf{H}}_{\text{p,LS}}) \right),
\end{equation}
where $\text{mat}(\cdot)$ and $\text{vec}(\cdot)$ denote the matricization and vectorization operators, $\nu$ represents SNR. $\mathbf{h}_\text{d}$ and $\mathbf{h}_\text{p}$ are the vectorized channel of data REs and pilot REs, while $\mathbf{R}_{\mathbf{h}_\text{d},\mathbf{h}_\text{p}}$ is the normalized cross-correlation between $\{\mathbf{h}_\text{d}$, $\mathbf{h}_\text{p}\}$ and $\mathbf{R}_{\mathbf{h}_\text{p},\mathbf{h}_\text{p}}$ is the normalized auto-correlation matrix of $\mathbf{h}_\text{p}$.

In practical channel estimation modules, the correlation matrices $\{ \mathbf{R}_{\mathbf{h}_\text{d},\mathbf{h}_\text{p}}, \mathbf{R}_{\mathbf{h}_\text{p},\mathbf{h}_\text{p}} \}$ can be calculated based on the prior knowledge of second-order channel statistics. Specifically, assuming a wide-sense stationary uncorrelated scattering (WSSUS) channel model, the normalized correlation between two REs separated by $\Delta k$ subcarriers and $\Delta n$ OFDM symbols can be decoupled into independent frequency- and time-domain components
\begin{equation} \label{eq:RH_sep_short}
R(\Delta k,\Delta n)=R_{\text{f}}(\Delta k)\,R_{\text{t}}(\Delta n),
\end{equation}
where $R_{\text{f}}(\Delta k)$ and $R_{\text{t}}(\Delta n)$ denote the normalized frequency- and time-domain correlations, respectively. These single-domain correlations are commonly constructed using parameterized models. For an efficient implementation, the work \cite{fd} considers a uniform power delay profile (PDP) with mean delay $\tau_{\mu}$ and delay width $\tau_{\mathrm{w}}$, thus the frequency correlation is given by
\begin{equation} \label{eq:RH_f_short}
R_{\text{f}}(\Delta k)=e^{-j2\pi \tau_{\mu}\Delta k\Delta f}\,
\text{sinc}\!\left(\pi\tau_{\text{w}}\Delta k\Delta f\right),
\end{equation}
where $\Delta f$ is the subcarrier spacing. Moreover, under the classical Jakes model with maximum Doppler frequency $f_\text{D}$, the time correlation is calculated as \cite{td}
\begin{equation} \label{eq:RH_t_short}
R_{\text{t}}(\Delta n)=J_0\!\left(2\pi f_\text{D} T_\text{s} \Delta n\right),
\end{equation}
where $T_\text{s}$ denotes the duration of one OFDM symbol, and $J_0(\cdot)$ is the zeroth-order Bessel function. 

In modern wireless systems, the channel parameters $\{\tau_{\mu},\tau_{\text{w}},f_{\text{D}},\nu\}$ can be estimated using maximum-likelihood methods \cite{td, fd} from semi-static reference signals (e.g., TRS in NR \cite{3gpp1}). Such reference signals are not subject to precoding and provide wideband observations, enabling reliable characterization of time-frequency channel correlations and effective Wiener filtering.

\section{Hypernetwork-Based UNet \\ for Channel Estimation}\label{S3}

\subsection{Problem Formulation}

Most existing deep learning-based channel estimation methods rely solely on LS estimates obtained from DMRS as the network input \cite{channelnet, reesnet, ceunet, attenreesnet, sccenet}. Such approaches do not explicitly exploit channel parameter information provided by semi-static reference signals, and therefore exhibit limited adaptability to different time-frequency correlation conditions.

In this letter, we consider a parameter-aware channel estimation framework by extending the network input beyond conventional pilot LS estimates. Concretely, in addition to the LS estimates at DMRS positions $\hat{\mathbf{H}}_{\text{p,LS}}$, a set of channel parameters measured from semi-static reference signals $\mathcal{P}$ are introduced as the auxiliary input, including the mean delay $\tau_\mu$, the delay width $\tau_\text{w}$, the maximum Doppler $f_\text{D}$, and the SNR $\nu$, i.e., $\mathcal{P} = \{\tau_\mu, \tau_\text{w}, f_\text{D}, \nu\}$. These parameters jointly characterize the frequency selectivity, temporal selectivity, and noise level of the wireless channel, which provide a compact description of the underlying channel correlation structure.
Therefore, the channel estimation problem is formulated as learning a mapping as follows
\begin{equation}
\mathbf{H} = g(\hat{\mathbf{H}}_{\text{p,LS}},\mathcal{P}; \Theta),
\end{equation}
where $g(\cdot)$ denotes the estimator, and $\Theta$ represents the corresponding model parameters.

\subsection{HyperCEUNet Design}

\begin{figure}[t] 
\begin{center}
\vspace{0mm}
\includegraphics[width=0.5\textwidth]{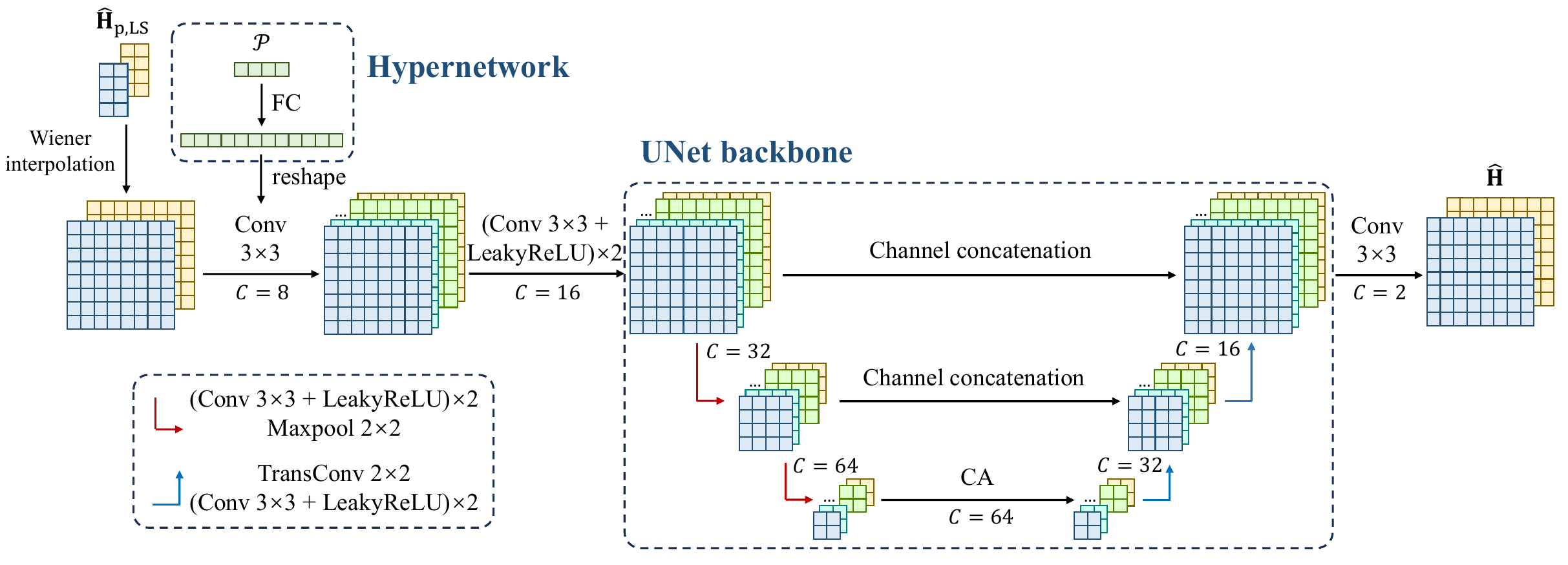}
\end{center}
\vspace{-4mm}
\caption{Illustration of proposed HyperCEUNet. $C$ denotes output feature channels of hidden layers. Conv, TransConv and FC denote convolutional layer, transposed convolutional layer and fully-connected layer, respectively. CA represents channel attention.}
\vspace{-2mm}
\label{fig1}
\end{figure}

To fully exploit the estimated channel parameters, we propose HyperCEUNet, a parameter-aware neural network architecture that modulates the UNet-based backbone via a hypernetwork, as shown in Fig.~\ref{fig1}. By conditioning the UNet feature extractor on channel parameter information, HyperCEUNet is able to adapt to different time-frequency correlation conditions with limited additional complexity. The overall design is detailed as follows.

\begin{figure}[t!] 
\begin{center}
\includegraphics[width=.4\textwidth]{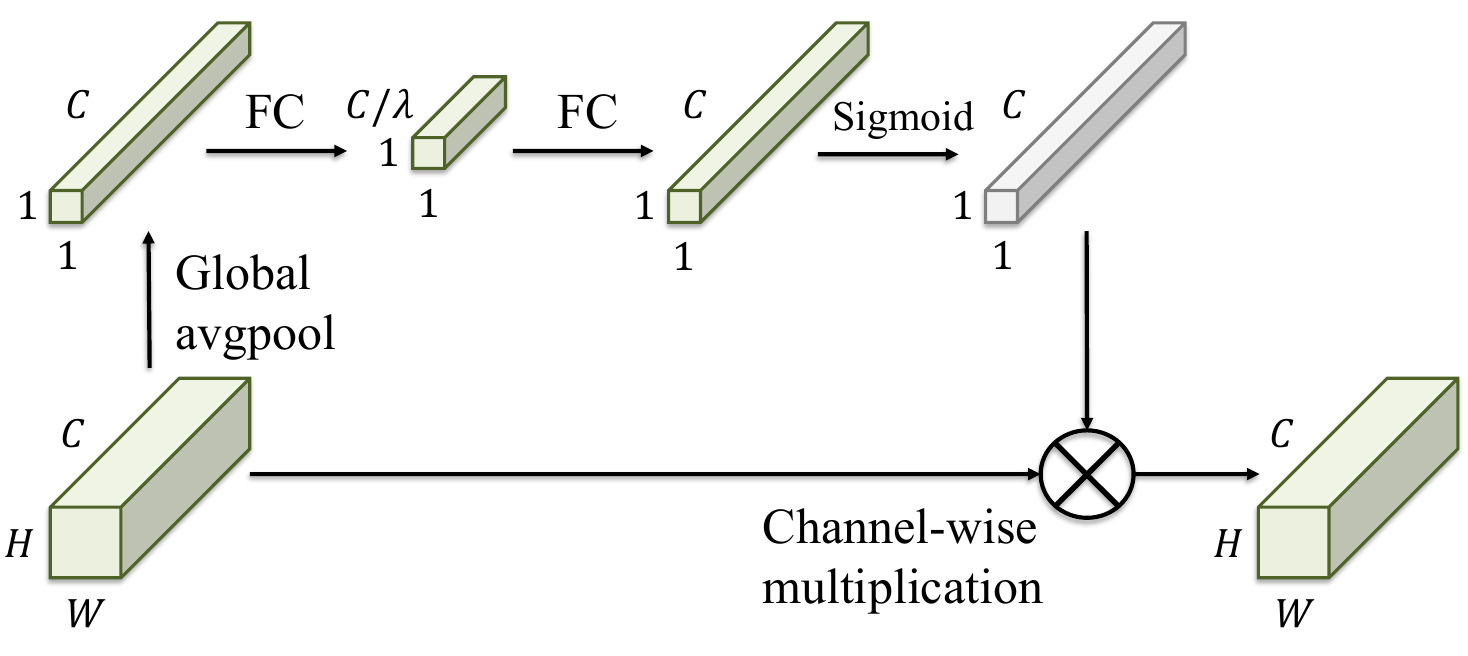}
\end{center}
\vspace{-4mm}
\caption{Illustration of CA module, where $C, H, W$ denote feature channels, height and width of input feature map, respectively.}
\vspace*{-4mm}
\label{fig2}
\end{figure}

\subsubsection{Wiener-Based Interpolation Enhancement}

Similar to existing works that formulate channel estimation as an image super-resolution problem \cite{channelnet, ceunet}, the LS estimates at DMRS positions are first interpolated to the entire time-frequency resource grid and used as the input to the neural network. Conventional approaches typically employ simple interpolation methods, such as bilinear interpolation \cite{channelnet, ceunet}, to fill data positions, which fail to effectively exploit the underlying channel correlation structure and often result in suboptimal performance.

In contrast, when channel parameters are available, the Wiener filter provides a principled way to estimate the channel at data positions by explicitly modeling the time-frequency correlation and noise statistics. Motivated by this observation, we adopt Wiener interpolation in (\ref{eq4}) to obtain an initial channel estimate at data positions, which serves as a correlation-aware initialization for the subsequent neural network. By incorporating this parameter-informed prior, the learning burden of neural network is significantly reduced, allowing it to focus on residual refinement rather than learning sophisticated long-range interpolation from scratch.

\subsubsection{UNet Backbone with CA}

UNet has demonstrated strong capability in capturing multi-scale features through an encoder-decoder structure with skip connections, and has been successfully applied to various image reconstruction tasks \cite{unet}. In our architecture, UNet is adopted as the backbone network to extract hierarchical representations of the channel over the time-frequency resource grid.

To enhance the representation capability of the UNet backbone, we introduce an attention mechanism at the bottleneck stage. Notably, after multiple downsampling operations in the encoder of UNet, the bottleneck feature map has a very low spatial resolution, which inherently limits the effectiveness of spatial attention. In contrast, the bottleneck typically contains a large number of feature channels, where redundancy becomes prominent and should be suppressed.
Motivated by this observation, instead of adopting attention mechanisms that jointly model spatial and channel dimensions \cite{ceunet}, we focus on CA \cite{ca} to recalibrate channel-wise features with negligible computational overhead, as shown in Fig.~\ref{fig2}. Specifically, let the bottleneck feature map be denoted by $\mathbf{F}\in\mathbb{R}^{C\times H\times W}$ with channel number $C$, height $H$, and width $W$. CA first performs global average pooling to obtain a channel descriptor $\mathbf{z}\in\mathbb{R}^{C}$, whose element $z_c=\frac{1}{HW}\sum_{i=1}^{H}\sum_{j=1}^{W}\mathbf{F}[c,i,j]$. The descriptor is then passed through a squeeze-and-excitation block to generate channel-wise attention weights as follows
\begin{equation}
\mathbf{a}=\sigma\!\left(\mathbf{W}_2\,\delta(\mathbf{W}_1\mathbf{z})\right),
\end{equation}
where $\mathbf{W}_1$ and $\mathbf{W}_2$ denote two fully-connected (FC) layers with a reduction ratio $\lambda$, $\delta(\cdot)$ is the ReLU activation, and $\sigma(\cdot)$ is the sigmoid function. Finally, the recalibrated feature map is obtained by channel-wise multiplication.
In addition, dropout is applied along the channel dimension at the bottleneck to alleviate overfitting caused by excessive redundancy.

\subsubsection{Hypernetwork-Based Pre-Filtering}
{\color{black}To enable parameter-aware adaptation, we introduce a hypernetwork \cite{hyper} that maps the estimated channel parameter set $\mathcal{P}$ to the weights of the pre-filtering layer. Since one network generates the parameters of another network, the proposed design is termed \emph{HyperCEUNet}.}
Directly applying hypernetwork-based modulation to all UNet parameters would incur excessive complexity due to the large parameter count. Therefore, we propose a lightweight alternative by inserting a single convolutional pre-filtering layer before the UNet backbone and modulating only this layer using the hypernetwork.
Specifically, the hypernetwork maps $\mathcal{P}$ to the convolutional kernel weights of the pre-filtering layer through a simple FC transformation as follows
\begin{equation}
\mathbf{w}_{\text{hy}} = g_\text{hy}(\mathcal{P}),
\end{equation}
where $g_\text{hy}(\cdot)$ denotes the hypernetwork and $\mathbf{w}_{\text{hy}}$ represents the generated convolutional parameters.

\begin{figure*}[t] 
\color{black}
\begin{center}
\includegraphics[width=0.72\textwidth]{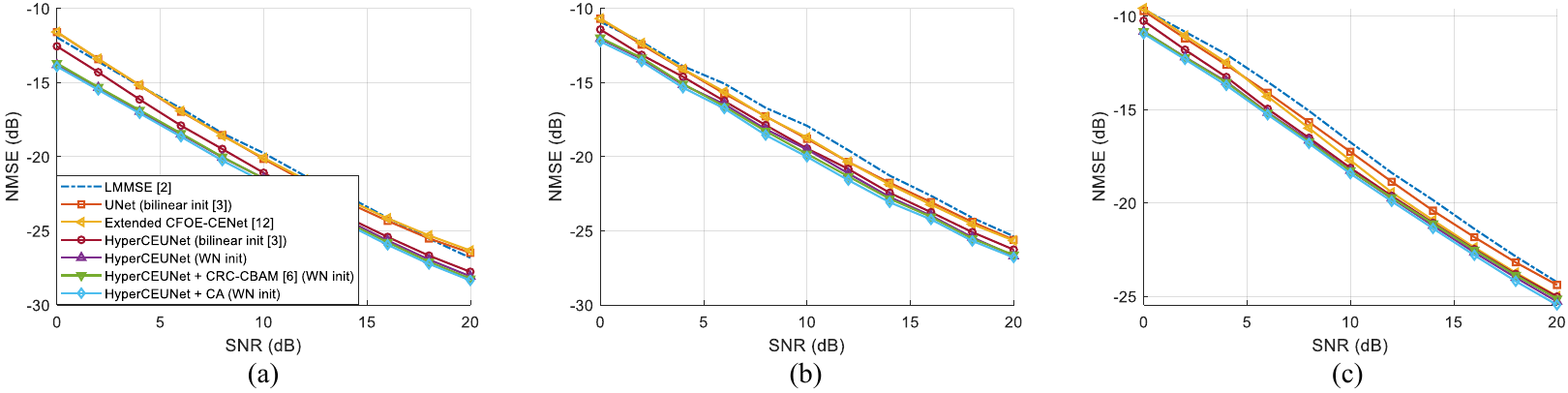}
\end{center}
\vspace{-6mm}
\caption{NMSE performance versus SNR under different channel models: (a) TDL-A, (b) TDL-B, and (c) TDL-C.}
\vspace{-4mm}
\label{fig3}
\end{figure*}

The resulting pre-filtering layer performs parameter-conditioned feature filtering before the UNet backbone, serving as a data-driven generalization of the Wiener filter.
By explicitly conditioning the filter behavior on channel parameters, this layer adapts the input features to different time-frequency correlation structures, while keeping the overall computational overhead low.

\subsubsection{Model Training}

The proposed HyperCEUNet is trained in a supervised manner using the mean squared error (MSE) loss between the estimated channel $\hat{\mathbf{H}}$ and the ground-truth channel $\mathbf{H}$. Specifically, for a mini-batch of $B$ training samples, the loss is computed as
\begin{equation}
\mathcal{L}=\frac{1}{B}\sum_{b=1}^{B}\left\|\hat{\mathbf{H}}^{(b)}-\mathbf{H}^{(b)}\right\|_\text{F}^2,
\end{equation}
where the superscript $(b)$ denotes the sample index.
{\color{black}It is noted that HyperCEUNet is adaptively modulated by the channel parameters associated with the current sample. Therefore, when the underlying channel parameter space is sufficiently covered during training, the proposed parameter-aware design is expected to generalize to unseen channel models as well.}

{\color{black}During inference, channel parameters are first estimated from semi-static reference signals, while LS estimates are obtained from DMRS. The estimated parameters are then used for Wiener-based initialization and hypernetwork-driven pre-filtering, after which the UNet backbone outputs the final channel estimate.}

{\color{black}Besides, the present HyperCEUNet design focuses on the TRS-configured scenario, where the parameter set $\mathcal{P}$ is assumed to be available through an auxiliary parameter estimation stage. Once $\mathcal{P}$ is given, a conventional LMMSE estimator can be used as a correlation-aware initialization, and HyperCEUNet further refines the channel estimate by compensating for residual errors. When TRS is unavailable, a residual hypernetwork design may be considered, where the pre-filtering layer keeps default weights and the hypernetwork only generates residual updates when TRS-assisted channel parameters are available.}

{\color{black}Although the single-input single-output (SISO) case is considered in this letter, the proposed hypernetwork-based parameter incorporation framework can be directly extended to MIMO scenarios. Specifically, the transmit and receive antenna dimensions can be folded into the DMRS-based LS input, while a 3D-convolution-based UNet backbone can be adopted to jointly exploit spatial and time-frequency correlations.
In addition, the current UNet backbone may be replaced by lightweight Transformer-based architectures \cite{transformer} in future work, to further reduce inference complexity and improve practical deployment efficiency.}

\section{ Simulation Analysis}\label{S4}

\subsection{Simulation Setup}\label{S4.1}

We evaluate HyperCEUNet over the Third Generation Partnership Project (3GPP) TDL-A/B/C channel models \cite{tdl} with maximum Doppler frequencies $f_{\text{D}}\in\{5,100,300\}$~Hz. The SNR $\nu$ ranges from $0$ to $20$~dB with a step size of $2$~dB. The considered time-frequency resource grid consists of $N_{\text{f}}=48$ subcarriers and $N_{\text{t}}=12$ OFDM symbols (i.e., $576$ REs in total). DMRS occupies $N_{\text{t,p}}=2$ OFDM symbols indexed as $\{0,9\}$ and $N_{\text{f,p}}=24$ subcarriers located at even indices $\{0,2,4,\ldots\}$. TRS occupies $4$ OFDM symbols indexed as $\{6,10,20,24\}$ (i.e., TRS is located in two consecutive subframes) and spans the entire system bandwidth containing $624$ subcarriers, where one TRS RE is mapped every four subcarriers in the frequency domain. Moreover, TRS is configured with a periodicity of $20$ subframes, providing wideband observations for reliable parameter estimation.

A mixed dataset containing $432{,}432$ samples is generated by combining all channel models, Doppler frequencies, and SNR conditions, with $80\%$ and $20\%$ used for training and validation, respectively. In contrast, testing is performed on separate datasets for each scenario, with $108{,}405$ testing samples in total. The channel parameters $\mathcal{P}$ are estimated using maximum-likelihood methods \cite{td, fd}. Each parameter is normalized independently before being fed into the hypernetwork, and the estimation errors are inherently included in the input to reflect practical conditions.

Regarding the HyperCEUNet configuration, most architectural details are illustrated in Fig.~\ref{fig1}. Specifically, the UNet backbone employs two-stage downsampling/upsampling with a base feature channel number of $16$, and the output feature channel number of our proposed pre-filtering layer is set to $8$. CA adopts a reduction ratio of $\lambda=8$, and channel-wise dropout with probability $0.3$ is applied at the bottleneck. All $3\times3$ convolutional layers use zero-padding of size $1$ and stride $1$, while all transposed convolutional layers adopt kernel size $2$ and stride $2$. Moreover, HyperCEUNet is trained using the Adam optimizer with a learning rate of $10^{-3}$.
{\color{black}Table~\ref{tab1} summarizes the trainable parameter counts of the proposed HyperCEUNet. Compared with the UNet backbone, the proposed hypernetwork-based pre-filtering introduces only $1{,}584$ additional trainable parameters, corresponding to a relative increase of $1.35\%$.}

\begin{table}[t]
\color{black}
\caption{Trainable parameter counts of HyperCEUNet.}
\label{tab1}
\centering
\begin{tabular}{cc}
\hline
Module & Trainable parameters \\
\hline
UNet backbone (w/o pre-filtering and CA) & 117,170 \\
Hypernetwork-based pre-filtering & 1,584 \\
CA & 1,096 \\
\hline
\end{tabular}
\vspace{-4mm}
\end{table}

\begin{figure*}[tp!]
\color{black}
\begin{center}
\vspace{0mm}
\includegraphics[width=0.68\textwidth]{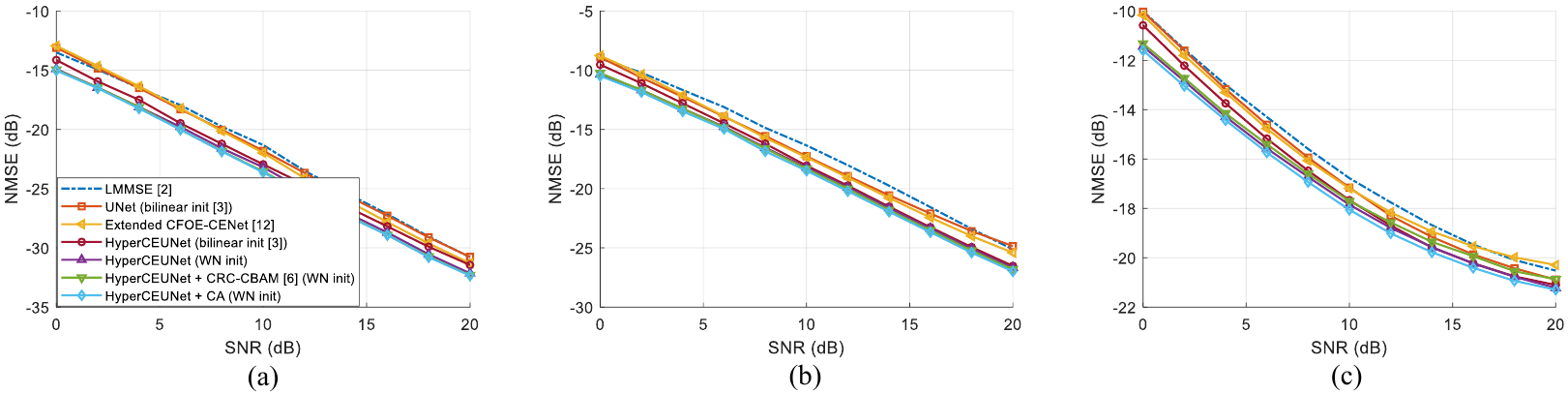}
\end{center}
\vspace{-6mm}
\caption{NMSE performance versus SNR under different Doppler frequencies: (a) $5$~Hz, (b) $100$~Hz, and (c) $300$~Hz.}
\vspace{-2mm}
\label{fig4}
\end{figure*}

\begin{figure*}[tp!]
\color{black}
\begin{center}
\vspace{0mm}
\includegraphics[width=0.68\textwidth]{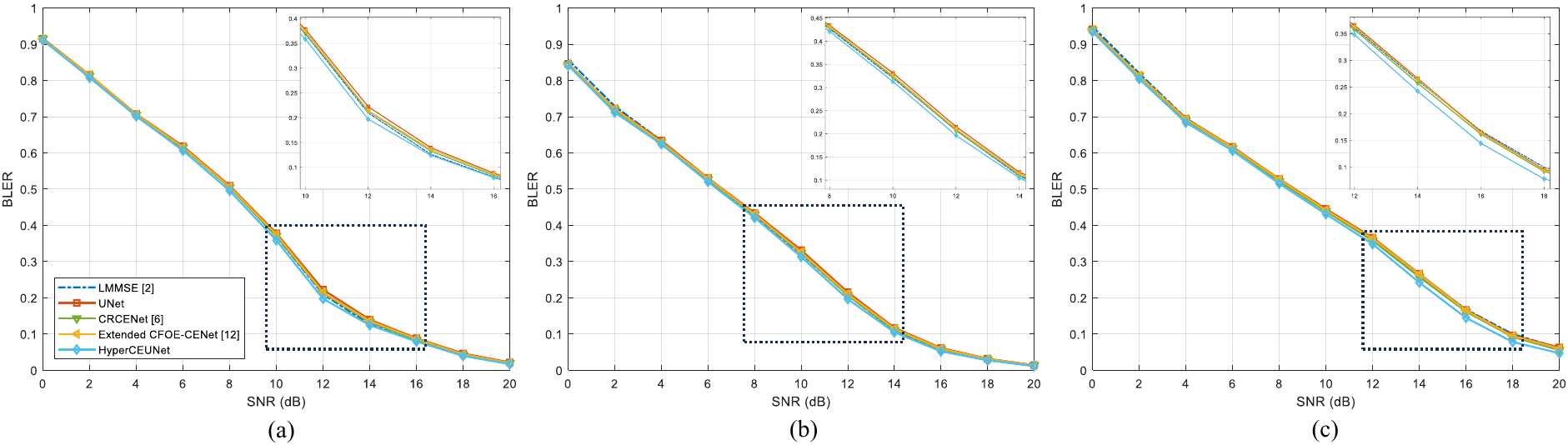}
\end{center}
\vspace{-6mm}
\caption{BLER performance versus SNR under different Doppler frequencies: (a) $5$~Hz, (b) $100$~Hz, and (c) $300$~Hz.}
\vspace{-4mm}
\label{fig5}
\end{figure*}

\subsection{Simulation Results}\label{S4.2}

The normalized mean squared error (NMSE) is adopted as the performance metric, and the following baselines are evaluated.
\emph{LMMSE} denotes the conventional Wiener filter-based LMMSE estimator \cite{lmmse} constructed from the estimated channel parameters.
\emph{UNet (bilinear init)} adopts the UNet backbone and applies bilinear interpolation \cite{channelnet} to initialize data resources. \emph{HyperCEUNet (bilinear init)} further introduces the proposed hypernetwork-based pre-filtering layer. \emph{HyperCEUNet (WN init)} replaces bilinear interpolation with Wiener interpolation based on the estimated channel parameters.
{\color{black}For a fair comparison, an extended variant of \emph{CFOE-CENet} \cite{joint2} is considered, where the auxiliary branch is designed to infer side information related to time-frequency selectivity and SNR from DMRS.}
In addition, two attention-enhanced variants are evaluated at the UNet bottleneck. \emph{HyperCEUNet + CRC-CBAM (WN init)} employs the Conv-ReLU-Conv convolutional block attention module (CRC-CBAM) {\color{black}proposed in CRCENet \cite{ceunet}}, whereas \emph{HyperCEUNet + CA (WN init)} adopts the CA module with negligible complexity.

Figs.~3 and 4 compare the NMSE performance of different methods under various channel models and Doppler frequencies. As observed, our proposed HyperCEUNet with CA and Wiener initialization consistently achieves the best NMSE across different scenarios over the entire SNR range. Compared with the UNet backbone, the proposed hypernetwork architecture provides clear performance improvement, demonstrating the effectiveness of incorporating parameter-aware adaptation into deep learning-based channel estimation. Besides, HyperCEUNet with Wiener initialization outperforms its bilinear counterpart, indicating that Wiener initialization is beneficial for sufficiently exploiting the underlying channel correlation structure. Moreover, incorporating CA further improves the NMSE performance compared to CRC-CBAM.

{\color{black}To further evaluate communication-level performance, coded block error rate (BLER) under different Doppler frequencies is compared in Fig.~5 for a transmission duration of $500$~ms, where NR modulation and coding scheme (MCS) $11$ with 64QAM modulation and target code rate $466/1024$ is adopted. As observed, the proposed HyperCEUNet achieves the best BLER performance, with more pronounced gains at high Doppler frequency.}

\vspace{-2mm}
\section{Conclusions}\label{S5}

This letter proposes HyperCEUNet, a parameter-aware deep learning framework for channel estimation that explicitly exploits channel parameters to enhance estimation accuracy. By leveraging semi-static reference signals for parameter estimation, a lightweight hypernetwork generates a parameter-conditioned pre-filtering layer before the CA-enhanced UNet backbone, and Wiener-based initialization provides a correlation-aware prior for data resource estimation. Simulation results demonstrate that HyperCEUNet achieves superior NMSE performance compared with
its conventional counterparts.

\vspace{-2.5mm}
{
}

\end{document}